\documentclass{article}%
\usepackage{amssymb}
\usepackage{amsmath}
\usepackage{revsymb}
\usepackage{eurosym}
\usepackage{graphicx}
\usepackage{amsfonts}%
\providecommand{\U}[1]{\protect\rule{.1in}{.1in}}
\def\TextSymbolUnavailable#1{\textbf{???}}
\ifx\pdfoutput\relax\let\pdfoutput=\undefined\fi
\newcount\msipdfoutput
\ifx\pdfoutput\undefined\else
\ifcase\pdfoutput\else
\ifx\paperwidth\undefined\else
\ifdim\paperheight=0pt\relax\else\pdfpageheight\paperheight\fi
\ifdim\paperwidth=0pt\relax\else\pdfpagewidth\paperwidth\fi
\fi\fi\fi
\begin{document}

\title{Integral Equation for CFT/string duality}
\author{A.A.Migdal}
\maketitle

\begin{abstract}
We reinterpret and extend some old work on CFT/string duality. We consider
some asymptotically conformal field theory in large N limit, with conformal
symmetry broken by VEV's of infinite number of operators. Assuming that this
theory confines (i.e. is dual to infinite number of free composite particles)
we derive explicit equation for the mass spectrum operator $Q$ of the theory,
relating this operator to terms OPE expansion of CFT. Under some general
assumptions about growth of OPE coefficients (less than double factorial
growth) the resulting expansion for the mass spectrum is convergent. This
method applies to confining CFT of ADS family as well as any asymptotically
CFT with confinement. This includes the ordinary QCD. In the latter case the
first terms of our perturbation expansion have good correspondence with
experimental Regge trajectories at low angular momentum.

\end{abstract}
\tableofcontents
\listoffigures

\section{Introduction}

Large N QCD (\cite{Hooft}) is a remarkable mathematical problem, having most
desirable universality and at the same time being close enough to reality.
Being solved with proper generality, it may serve as a basis for quantitative
theory of hadrons, by means of systematic expansion in inverse powers of
$N_{c}^{2}$. However, all attempts to do so in the past met the brick wall of
infrared divergencies.

In most simple terms, the infrared divergency is the fact that spectrum of the
theory is discrete whereas in perturbation theory it is continuous. There are
freely flying interacting quarks and gluons, which all must get confined so
that physical spectrum consists instead of infinitely rising Regge
trajectories of mesons and glueballs.%

%TCIMACRO{\FRAME{ftbpF}{5.361in}{4.0231in}{0in}{}{}{duality-plane.025.ps}%
%width 5.361in;  height 4.0231in;  depth 0in;  original-width 8.1699in;
%original-height 6.1203in;  cropleft "0";  croptop "1";  cropright "1";
%cropbottom "0";  filename 'duality-plane.025.ps';file-properties "XNPEU";}} }%
%BeginExpansion
\begin{figure}[ptb]%
\centering
\ifcase\msipdfoutput
\includegraphics[
height=4.0231in,
width=5.361in
]%
{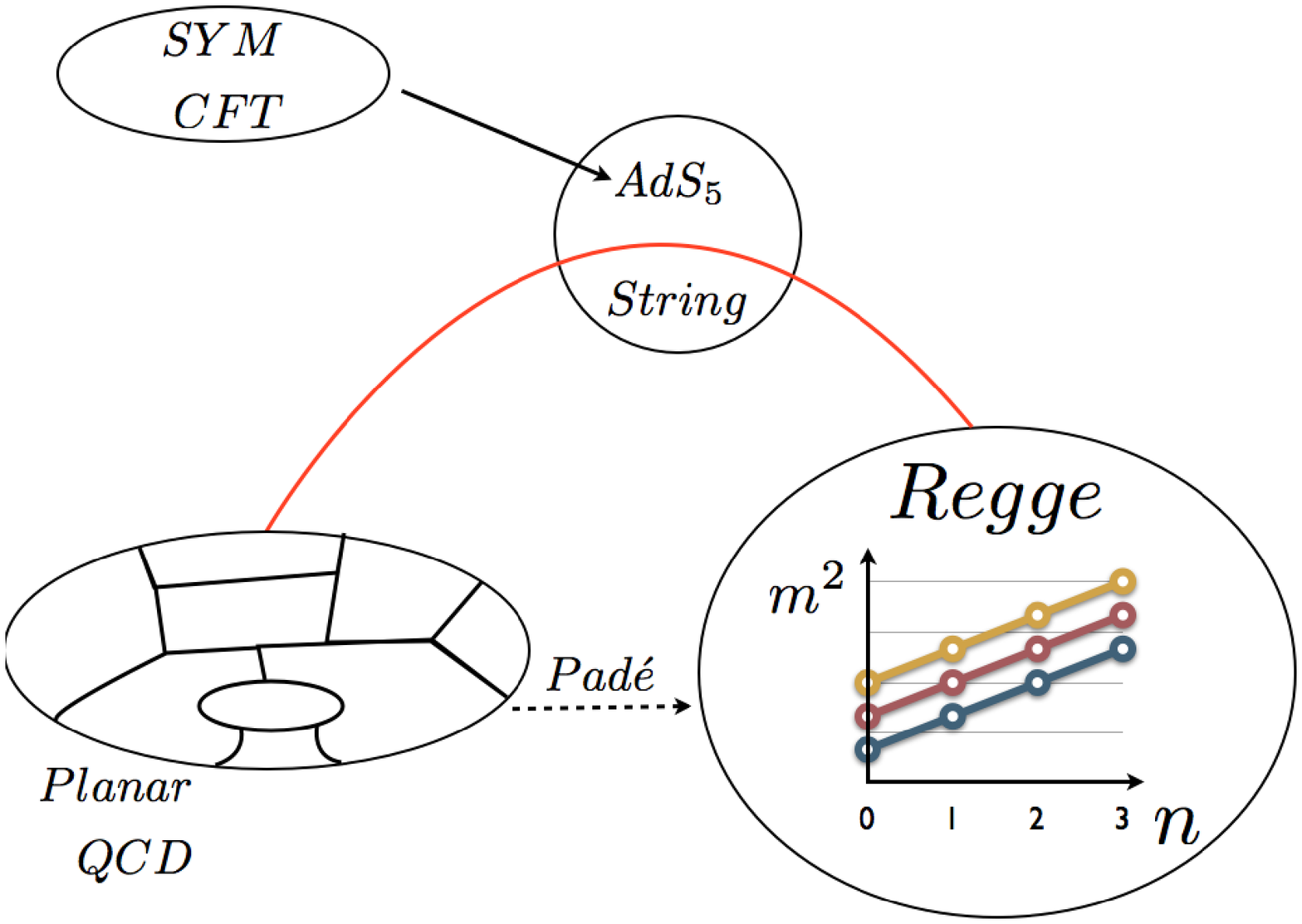}%
\else
\includegraphics[
height=4.0231in,
width=5.361in
]%
{C:/Users/Sasha/Documents/graphics/duality-plane.pdf}%
\fi
\end{figure}
%EndExpansion

Everybody believes in this picture now, though it was never rigorously proven.
The real question is how to transform the sum of planar graphs of Large N QCD
into some sensible expansion for hadronic spectrum. This must be doable, as
the effective coupling at hadron scale is not so large, and hadrons are not
far from some bag model of free quarks. The spatial bag is of course
unacceptable, as it totally screws the particle spectrum being
non-relativistic and transitionally non-invariant.

It was almost 35 years ago that I attempted to solve this problem and made
some advances (\cite{MigdalPade},\cite{MigdalAlpha}). I developed systematic
method of approximating the meromorphic function like 2-point function of
gauge invariant composite fields in Large N QCD as a sum of infinite number of
pole terms with positive residues. For such an approximation to be unique it
has to have higher than powerlike convergence at large Euclidean momenta --
otherwise moving poles or changing residues would be allowed without changing
leading powerlike term as long as the net change in power terms decreases
faster than this asymprotic term, say $\log(t)$ with $t=p^{2}$ being usual
momentum squared in Minkowski space.

For example the first meromorphic function which comes to the mind of any grad
student would be $\Psi(z)=\Gamma^{\prime}(z)/\Gamma(z),z=-t$ but it approaches
its logarithmic asymptotics with powerlike corrections which means that
positions and residues of its poles are ambiguous. On the contrary, the ratio
of Bessel functions approaches its logarithmic asymptotics with exponential
accuracy. The soft perturbations of CFT corresponding to negative power terms
in $p^{2}$ will be altered by powerlike terms like we have in the $\Psi$
function, but preserved with ratio of Bessel functions. Conversely, any such
soft correction must modify the Bessel functions to something else otherwise
it will never appear in asymptotic expansion. In this paper we will once again
explain how the perturbations of the mass spectrum \ can be recovered from the
soft perturbations of CFT.

There are theorems in Pade theory \cite{ZinnJustin} which guarantee that so
called Stieltjes-function (analytic in cut plane with positive discontinuity
across the cut) preserve this property in any order of approximation. The
poles of Pade approximant are all located along the cut with positive residues
-- so that discontinuity reduces to finite sum of positive $\delta$ terms.
This Stieltjes-function property generalizes to arbitrary matrix functions of
one variable in which case it becomes equivalent to one particle unitarity +
analyticity. In the limit when number of particles goes to infinity this
property is just what we need in Large N QCD.

So, the limit of infinite number of poles with fixed positions, introduced and
studied in \cite{MigdalPade} represents the correct framework for large N
confining theory regardless of its asymprotic properties at large Euclidean
momenta. In case of asymptotically conformal (or asymptotically free) theory
the Pade regularization further simplifies and produces explicit calculable
results depending of values of operator dimensions (normal or anomalous).

In this paper we briefly summarize this theory and advance it further,
producing explicit terms of perturbation expansion of QCD mass ratios in terms
of calculable terms of perturbation expansion of the (matrix of) 2-point
functions of conformal fields. We use dimensional regularization which fits
nicely into our framework, and supplement it by infrared regularization using
Matrix Pade theory in the limit of infinite number of poles. Important step is
that we are able to eliminate the infrared cutoff in every order in
dimensionless effective coupling $\alpha$, normalized so that it must be set
to $1$ after summation of perturbation expansion.

This is continuation of old work \cite{MigdalPade,MigdalAlpha}, but unlike
that old work, now we produce analytic rather than numerical formulas. Given
planar graphs for the matrix of 2-point function, which are universal
functions of $\epsilon$ times powers of $\lambda t^{-\frac{\epsilon}{2}}$where
$\lambda$ is t'Hoofts coupling constant and $d=4-\epsilon$ is dimension of
space. Note that at any positive $\epsilon<3$ we still expect confinement to
hold, as it is known for $\epsilon=1,2$. The spectral function of 2-point
functions of renormalized conformal tensor fields in QCD are finite in any
order of expansion in running coupling constant, which we normalize at the
infrared cutoff $R$.

Yes, we are simply presenting infinity-free perturbation expansion for Large N
QCD observables with calculable coefficients and nice physical properties
(zeroth approximation corresponds to masses proportional to roots of Bessel
functions in agreement with experiment). Terms of our expansion correspond to
planar graphs, but after Pade regularization and dimensional transmutations
all cutoffs disappear and we get universal numbers just as we should. There is
no ambiguity nor approximations in computing these numbers, do not get misled
by the words "Pade approximation". We start with approximation, but later take
the limit returning us to original theory after renormalizing coupling
constant. The infrared cutoff $R$ describing spacial scale of the infrared
regularization is tend to $\infty$ term by term in perturbation expansion (to
be more precise this limit corresponds to effective coupling constant
$\alpha\rightarrow1$).

\section{Matrix Pade approximation in Hilbert Space}

\bigskip%
%TCIMACRO{\FRAME{ftbpF}{5.361in}{4.0231in}{0in}{}{}{complex-plane.004.ps}%
%width 5.361in;  height 4.0231in;  depth 0in;  original-width 8.1699in;
%original-height 6.1203in;  cropleft "0";  croptop "1";  cropright "1";
%cropbottom "0";  filename 'complex-plane.004.ps';file-properties "XNPEU";}} }%
%BeginExpansion
\begin{figure}[ptb]%
\centering
\ifcase\msipdfoutput
\includegraphics[
height=4.0231in,
width=5.361in
]%
{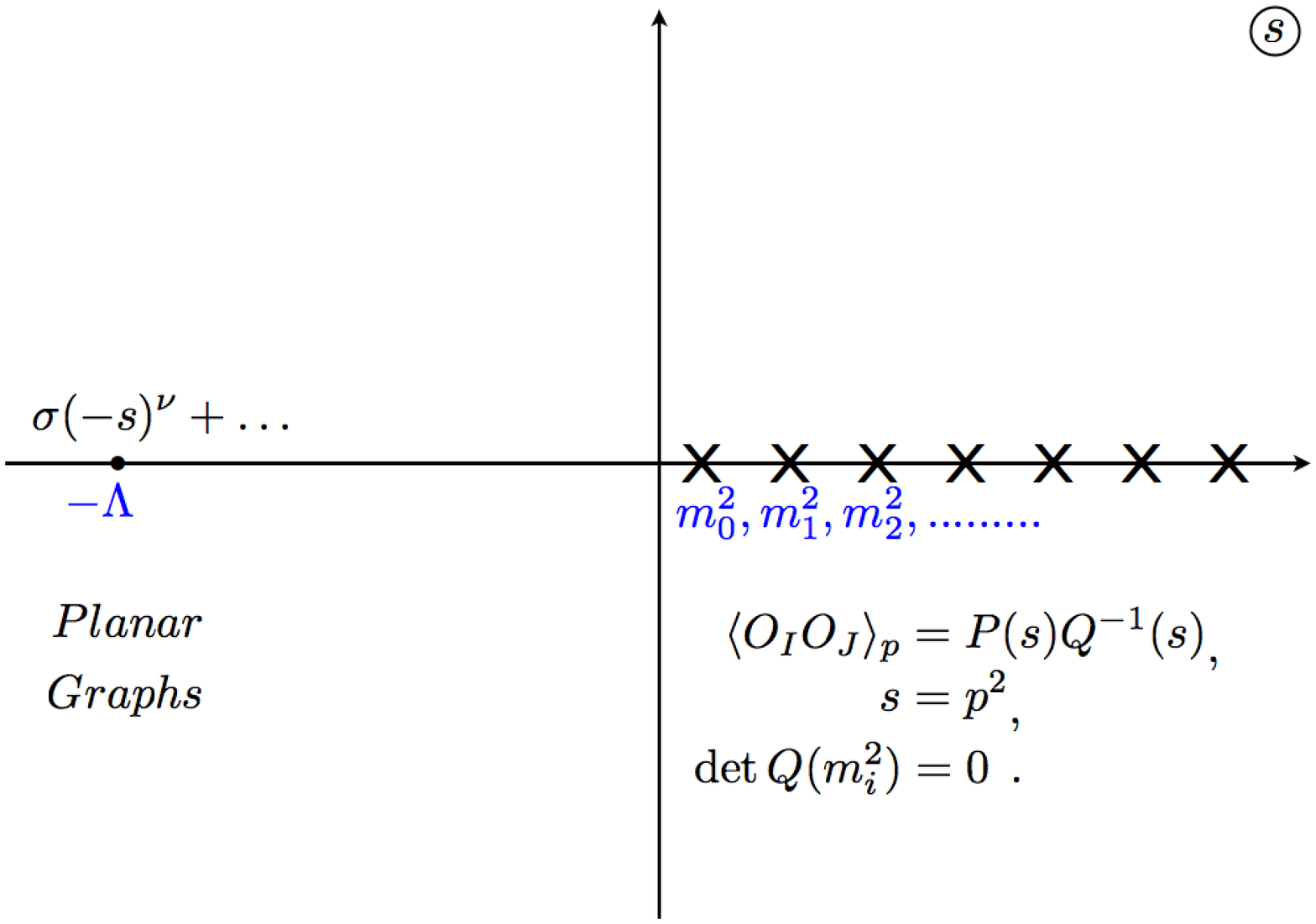}%
\else
\includegraphics[
height=4.0231in,
width=5.361in
]%
{C:/Users/Sasha/Documents/graphics/complex-plane.pdf}%
\fi
\end{figure}
%EndExpansion

Let us consider some CFT perturbed by some set of soft operators such as mass
terms. We shall further assume that this is confining $N_{c}=\infty$ theory,
with only planar graphs left and discrete spectrum of masses rising all the
way to infinity to match CFT asymptotics in deep Euclidean region of momenta.
In this limit we know very important property of the infinite matrix $G_{IJ}$
of 2-point functions of bilinear quark operators $\hat{O}_{J}(x)$: free
particle unitarity + analyticity. This is meromorphic matrix function of the
form:%
\begin{equation}
G_{IJ}(p^{2})=\int d^{d}xe^{ipx}\left\langle \hat{O}_{I}(0),\hat{O}%
_{J}(x)\right\rangle =\sum_{i}\frac{Z_{i}}{m_{i}^{2}-p^{2}}\Psi_{I}^{i}%
\Psi_{J}^{\dagger i}. \label{TwoPointFunction}%
\end{equation}
We are working with functions of single variable $p^{2}$ (in Minkowski metric)
assuming that kinematical factors depending on direction of momentum are
involved in definition of states $I,J$. So, in general the state $\left\vert
I\right\rangle $ depends upon $n_{\mu}=\frac{p_{\mu}}{\left\vert p\right\vert
}$ $\ $and the 2-point function $G_{IJ}(p^{2})$ is an irreducible tensor built
of $n_{\mu}$. For example, for conserved vector currents $\hat{O}%
_{J}=\overline{\psi}\gamma_{\mu}\psi$ there must be tensor $\delta_{\mu\nu
}-n_{\mu}n_{\nu}$. In higher order operators $\overline{\psi}\Gamma_{A}%
\nabla_{\alpha_{1}}\nabla_{\alpha2}...\nabla_{\alpha_{n}}\psi$ with
$\Gamma_{A}$ being one of 16 independent matrices for Dirac spinors, there
will be multiple invariant tensor terms in 2-point function, depending
on\ $n_{\mu}$. We shall ignore these details at this general stage of
discussion, leaving them for the next Section where we switch to QCD.
Unitarity implies that matrix spectral density%
\begin{equation}
\rho(t)=\Im G(t+i0),
\end{equation}
vanishes outside of the positive axis where it reduces to infinite number of
positive definite pole terms%
\begin{equation}
\rho(t)=\pi\sum_{i}Z_{i}\delta\left(  t-m_{i}^{2}\right)  \Psi^{i}\otimes
\Psi^{\dagger i},Z_{i}>0,m_{i}^{2}>=0.
\end{equation}
The matrix positivity property of the spectral function%
\begin{equation}
\left\langle a\left\vert \rho(t)\right\vert a\right\rangle >=0,
\end{equation}
for arbitrary state $\left\vert a\right\rangle $ as well as absence of
singularities outside positive axis in complex $t$ plane is characteristic of
so called Stieltjes functions. The discrete spectrum is not necessary for a
function to belong to Stieltjes class, for example $\sqrt{-t}$ is a Stieltjes
function. More up to a point, the 2-point function in any order of naive
perturbation expansion is also a Stieltjes function. We expect it to remain
such function beyond perturbation theory, but the continuum spectrum condense
to a discrete one.

In the leading order of any CFT, including of course the free quark-gluon
theory, the operators of different dimension do not correlate. There are
growing conformal families of operators with the same dimension, corresponding
to eigenstates of dilatation operator. In general case, beyond leading
approximation, the matrix of 2-point functions will become non-diagonal, so
that more general version of Pade theory must be employed to guarantee
unitarity + analyticity in whole Hilbert space of infinite number of free mesons.

\bigskip

This theory (\cite{ZinnJustin}) is a straightforward matrix generalization of
one-dimensional Pade theory. The Pade approximant is essentially a continued
fraction, which is obtained by sequence of transformations%
\begin{align}
G_{n}(\xi)  &  =A_{n}\frac{1}{1-\xi G_{n+1}(\xi)}A_{n}^{\dagger}%
,\label{ContFrac}\\
\xi &  =\left(  1+\frac{t}{\Lambda}\right)  ,\\
G(t)  &  =G_{0}\left(  \xi\right)  ,\\
A_{n}A_{n}^{\dagger}  &  =G_{n}(0);\nonumber
\end{align}
The constant matrix $A_{n}$ is defined up to irrelevant right multiplication
by unitary matrix, for a positive Hermitian $G_{n}(0)$ in euclidean region
$t=-\Lambda$ in terms of its eigenvectors $\left\langle a\left\vert i\right.
\right\rangle $and positive eigenvalues $g_{i}$%
\begin{align}
\left\langle a\left\vert G_{n}(0)\right\vert b\right\rangle  &  =\sum
_{i}\left\langle a\left\vert i\right.  \right\rangle g_{i}\left\langle
i\left\vert b\right.  \right\rangle ,\\
\left\langle a\left\vert A_{n}\right\vert i\right\rangle  &  =\left\langle
a\left\vert i\right.  \right\rangle \sqrt{g_{i}}.
\end{align}

Iterating these transformations forward we obtain the continued fraction%
\begin{equation}
G_{0}\left(  \xi\right)  =A_{0}\frac{1}{1-\xi A_{1}\frac{1}{1-\xi A_{2}%
\frac{1}{1-...}A_{2}^{\dagger}}A_{1}^{\dagger}}A_{0}^{\dagger};
\end{equation}
\qquad

Let us invert \ref{ContFrac}%
\begin{equation}
G_{n+1}(\xi)=\frac{1}{\xi}\left(  1-A_{n}^{\dagger}\frac{1}{G_{n}(\xi)}%
A_{n}\right)  . \label{InverseContFrac}%
\end{equation}

In order to find the value of $G_{n+1}(0)$ we need to expand $G_{n}%
(\xi)\rightarrow G_{n}(0)+\xi G_{n}^{^{\prime}}(0)$ and we find%
\begin{equation}
G_{n+1}(0)=A_{n}^{\dagger}\frac{1}{G_{n}(0)}G_{n}^{^{\prime}}(0)\frac{1}%
{G_{n}(0)}A_{n},
\end{equation}

This way we can recursively find these coefficients uniquely from the Taylor
expansion of original function $G(t)$ $=G_{0}\left(  \xi\right)  $.

The continued fraction will have $\frac{P}{Q}$ form with polynomial numerator
and denominator determined from recurrent equations%
\begin{align}
G_{n}\left(  \xi\right)   &  =P_{n}(\xi)\frac{1}{Q_{n}(\xi)}%
,\label{PQRecurrentEqs}\\
A_{n}A_{n}^{\dagger}  &  =G_{n}(0);\\
Q_{n+1}(\xi)  &  =\frac{1}{A_{n}}P_{n}(\xi),\\
P_{n+1}(\xi)  &  =\frac{1}{A_{n}}\frac{P_{n}(\xi)-G_{n}(0)Q_{n}(\xi)}{\xi}%
\end{align}
Truncation of continued fraction to $\left[  \frac{N}{N}\right]  $ approximant
corresponds to setting%
\begin{align}
Q_{N+1}(\xi)  &  =1,\\
P_{N+1}(\xi)  &  =0
\end{align}

and iterating equations backwards to $P_{0},Q_{0}$ the same equation in its
inverse form:%
\begin{align*}
P_{n}(\xi)  &  =A_{n}Q_{n+1}(\xi),\\
Q_{n}(\xi)  &  =\frac{1}{G_{n}(0)}\left(  P_{n}(\xi)-\xi A_{n}P_{n+1}%
(\xi)\right)  .
\end{align*}
In the $\frac{P}{Q}$ form the conservation of positivity is not immediately
clear, but in original form \ref{ContFrac} we may prove it as follows. By
taking discontinuity at $\xi^{+}$ at positive real $\xi>1$ of inverse relation
\ref{InverseContFrac} we find%
\begin{align}
\Im G_{n+1}(\xi^{+})  &  =\frac{1}{\xi}A_{n}^{\dagger}\frac{1}{G_{n}(\xi^{-}%
)}\Im G_{n}(\xi^{+})\frac{1}{G_{n}(\xi^{+})}A_{n},\\
\left\langle \alpha\left\vert \Im G_{n+1}(\xi^{+})\right\vert \alpha
\right\rangle  &  =\left\langle \beta\left\vert \Im G_{n}(\xi^{+})\right\vert
\beta\right\rangle >=0,\\
\left\vert \beta\right\rangle  &  =\frac{1}{\sqrt{\xi}}\frac{1}{G_{n}(\xi
^{+})}A_{n}\left\vert \alpha\right\rangle
\end{align}

Thus, by induction, all matrix functions $G_{n}(\xi)$ starting with original
one at $n=0$ are Stieltjes matrix functions. Moreover, we can prove by
induction that all matrix functions $G_{n}(\xi)$ are Hermitian matrices at
real $\xi<1$, corresponding to Euclidean region of momenta. Takin Hermitian
conjugate of \ref{InverseContFrac} and assuming that $G_{n}(\xi)$ is Hermitian
we see that the same is true about $G_{n+1}(\xi)$:
\begin{align}
G_{n+1}^{\dagger}(\xi)  &  =\frac{1}{\xi}\left(  1-A_{n}^{\dagger}\frac
{1}{G_{n}^{\dagger}(\xi)}A_{n}\right) \\
&  =\frac{1}{\xi}\left(  1-A_{n}^{\dagger}\frac{1}{G_{n}(\xi)}A_{n}\right)
\nonumber\\
&  =G_{n+1}(\xi).\nonumber
\end{align}
\qquad

Note that there is kind of gauge invariance of Pade approximant with respect
to right matrix multiplication by a constant matrix $W$ (independent of $t)$:
\begin{align}
Q(t)  &  \rightarrow Q(t)W,\\
P(t)  &  \rightarrow P(t)W
\end{align}

As for the left matrix multiplication it results in similarity transformation
for $PQ^{-1}$ so we can use it to diagonalize matrix $G$ at normalization
point $t=-\Lambda$.We use the gauge invariance to choose convenient
normalization of these $P,Q$ below.

The continued fraction being good practical way to build finite order
approximant, their general properties in the limit of large $N$ are better
studied in the form suggested by Pade himself for the general $\left[
\frac{M}{N}\right]  $ approximant:%
\begin{equation}
G_{0}(\xi)Q_{0}(\xi)-P_{0}(\xi)=O(\xi^{N+M+1}). \label{GeneralPadeEq}%
\end{equation}

Comparing coefficients at $\xi^{r},r=0,1,...N+M$ we get system of linear
matrix equations for matrix coefficients of polynomials $P_{0}(\xi),Q_{0}%
(\xi)$ relating these coefficients to Taylor coefficients of $G_{0}(\xi)$.
Namely, the last $N-1$ equations, which do not involve $P_{0}$ (its expansion
ending at $\xi^{M}$) produce linear equations for coefficients of $Q_{0}$:%
\begin{equation}
\left[  G_{0}(\xi)Q_{0}(\xi)\right]  _{\{M+1,M+N\}}=0,
\end{equation}

where $\left[  ...\right]  _{\{n,m\}}$ stands for part of Taylor expansion
with degrees from $n$ to $m$.

Adding gauge condition, say $Q_{0}(0)=1$ we get unique solution for $Q_{0}$
after which $P_{0}$ can be obtained directly as%
\begin{equation}
P_{0}(\xi)=\left[  G_{0}(\xi)Q_{0}(\xi)\right]  _{\{0,M\}} \label{GQ_0M}%
\end{equation}

Note that in this method we did not have to take any square roots from
$G_{0}(\xi)$ . So, is this really the same solution as we have built above
using continued fraction? Let us check at the $\left[  \frac{0}{1}\right]  $
approximant. In continued fraction we get%
\begin{align}
G_{0}\left(  \xi\right)   &  =A_{0}\frac{1}{1-\xi A_{1}A_{1}^{\dagger}}%
A_{0}^{\dagger}\\
&  =A_{0}\frac{1}{1-\xi G_{1}(0)}A_{0}^{\dagger}\nonumber\\
&  =A_{0}\frac{1}{1-\xi A_{0}^{\dagger}\frac{1}{G_{0}(0)}G_{0}^{^{\prime}%
}(0)\frac{1}{G_{0}(0)}A_{0}}A_{0}^{\dagger}\nonumber\\
&  =\frac{1}{\frac{1}{A_{0}}-\xi A_{0}^{\dagger}\frac{1}{G_{0}(0)}%
G_{0}^{^{\prime}}(0)\frac{1}{G_{0}(0)}}A_{0}^{\dagger}\nonumber\\
&  =\frac{1}{\frac{1}{A_{0}A_{0}^{\dagger}}-\xi\frac{1}{G_{0}(0)}%
G_{0}^{^{\prime}}(0)\frac{1}{G_{0}(0)}}\nonumber\\
&  =\frac{1}{\frac{1}{G_{0}(0)}-\xi\frac{1}{G_{0}(0)}G_{0}^{^{\prime}}%
(0)\frac{1}{G_{0}(0)}}\nonumber\\
&  =G_{0}(0)\frac{1}{1-\xi\frac{1}{G_{0}(0)}G_{0}^{^{\prime}}(0)}\nonumber\\
&  =P\frac{1}{Q}\nonumber
\end{align}

which is the same we would have obtained much easier from matrix Pade
equations:%
\begin{equation}
\left(  G_{0}(0)+G_{0}^{^{\prime}}(0)\xi\right)  (1+q_{1}\xi)-p_{0}=O(\xi^{2})
\end{equation}

Note that we could have represented the same continuos fraction differently:%
\begin{align}
G_{0}\left(  \xi\right)   &  =A_{0}\frac{1}{1-\xi A_{1}A_{1}^{\dagger}}%
A_{0}^{\dagger}\\
&  =...\\
&  =\frac{1}{\frac{1}{G_{0}(0)}-\xi\frac{1}{G_{0}(0)}G_{0}^{^{\prime}}%
(0)\frac{1}{G_{0}(0)}}\\
&  =\frac{1}{1-\xi G_{0}^{^{\prime}}(0)\frac{1}{G_{0}(0)}}G_{0}(0)\\
&  =\frac{1}{\tilde{Q}}\tilde{P}.
\end{align}
\qquad\qquad

This is so called left matrix approximant, which is just another way to
represent the same continuos fraction. The difference arises at the moment of
truncation and tends to zero with order of approximation, as both approximants
converge to the Stiltjes matrix function with exponential or better
convergence rate. In case of meromorphic function under consideration the
convergence means that positions and residues of poles converge to correct
values. Interesting fact is that the Pade approximation for masses decrease
monotonously with order $N$ of approximation so they always overestimate true
mass spectrum and converge to every mass from above. This remarkable property
will be used below (see also \cite{MigdalPade}).

Writing Pade equation \ref{GeneralPadeEq} as dispersion integral with matrix
spectral density we find set of linear equations for $Q$
\begin{equation}
\int_{0}^{\infty}ds\left(  1+s/\Lambda\right)  ^{(r-2N-1)}\rho
(s)Q(s)=0;r=0,...N, \label{IntegralPadeEq}%
\end{equation}

These equations mean that $Q(s)$ is an orthogonal matrix polynomial with
respect to matrix measure%
\begin{equation}
d\sigma(s)=ds\left(  1+s/\Lambda\right)  ^{(-2N-1)}\rho(s)
\end{equation}

The general solution \ref{GQ_0M} for $P$ can be rewritten as dispersion integral%

\begin{align}
P(t)  &  =\left[  G_{0}(\xi)Q_{0}(\xi)\right]  _{\{0,N\}}\label{PSolution}\\
&  =G_{0}(\xi)Q_{0}(\xi)-\left[  G_{0}(\xi)Q_{0}(\xi)\right]  _{\{N+1,\infty
\}}\\
&  =G(t)Q(t)-\int_{0}^{\infty}\frac{ds}{\pi(s-t)}\left(  \frac{\left(
1+t/\Lambda\right)  }{(1+s/\Lambda)}\right)  ^{N+1}\rho(s)Q(s).
\end{align}

\section{Green's Function of Pade Equations}

\bigskip

Let us represent the matrix spectral density as sum of power terms with
decreasing powers, coming from soft perturbations of CFT:%

\begin{align}
\left\langle I\left\vert \rho(s)\right\vert J\right\rangle  &  =s^{\nu_{I}%
}\left(  \left\langle I\left\vert \sigma\right\vert J\right\rangle +\sum
_{K}\left\langle I\left\vert g_{K}\right\vert J\right\rangle s^{-\Delta
_{K}^{IJ}}\right)  ,\\
\nu_{I} &  =\Delta_{I}-\Delta_{I}^{0},\Delta_{I}^{0}=d-2+n,\\
\Delta_{K}^{IJ} &  =\left(  \Delta_{I}-\Delta_{J}+\mu_{K}\right)  /2;
\end{align}

Here $\mu_{K}>0$ are mass dimensions of operators $g_{K}$ . The normal
dimension $\Delta_{I}^{0}=d-2+n$ was subtracted from $\Delta_{I}$ to account
for kinematical factors such as $\delta_{\mu\nu}p^{2}-p_{\mu}p_{v}$ for the
tensor of rank $n$ . These tensors must be factored out before applying
Pad\'{e} transformation. The remaining partial amplitudes represent the
matrrix which we treat as meromorphic function of $p^{2}$. 

Expansion in $K$ goes in negative powers of $s$ corresponding to UV asymptotic
expansion. The matrix $\hat{\sigma}$ is independent of $s$ and is block
diagonal in space of conformal operators. It is some tensor made of $n_{\mu}$
with simple properties.

The linear integral equation for $Q$ can be solved exactly using Greens
function satisfying (\cite{MigdalPade}):%
\begin{equation}
\int_{0}^{\infty}dtt^{\nu}\left(  1+t/\Lambda\right)  ^{(r-2N-1)}G_{\nu
}(t,s)=s^{\nu}\left(  1+s/\Lambda\right)  ^{(r-2N-1)};r=0,...N.
\end{equation}

Replacing the factor
\begin{equation}
s^{\nu_{I}}\left(  1+s/\Lambda\right)  ^{(r-2N-1)}%
\end{equation}
\qquad\ 

in \ref{IntegralPadeEq} by
\begin{equation}
\int_{0}^{\infty}dtt^{\nu_{I}}\left(  1+t/\Lambda\right)  ^{(r-2N-1)}%
G_{\nu_{I}}(t,s)
\end{equation}

we find%

\begin{equation}
\int_{0}^{\infty}dsG_{\nu_{I}}(t,s)\sum_{J}\left(  \left\langle I\left\vert
\sigma\right\vert J\right\rangle +\sum_{K}\left\langle I\left\vert
g_{K}\right\vert J\right\rangle s^{-\Delta_{K}^{IJ}}\right)  \left\langle
J\left\vert Q(s)\right\vert M\right\rangle =S_{IM}(t),
\end{equation}

\begin{equation}
\int_{0}^{\infty}dt\left(  1+t/\Lambda\right)  ^{(r-2N-1)}t^{\nu_{I}}%
S_{IM}(t)=0,r=0,...N-1.
\end{equation}

We know solutions of above equations for $G_{\nu}(t,s)$ and $S_{IM}(t)$
(\cite{MigdalPade}):%
\begin{align}
S_{IM}(t) &  =\oint_{C}\frac{d\omega}{2\pi i}f_{\nu_{I}}(\omega)\left(
1+t/\Lambda\right)  ^{\omega}\left\langle I\left\vert \hat{W}\right\vert
M\right\rangle ,\\
f_{\nu}(\omega) &  =\frac{\Gamma(2N+1-\omega)\Gamma(-\omega)N^{2(1-\nu)}%
}{\Gamma(N+1-\nu-\omega)\Gamma(N+1-\omega)}.\\
G_{\nu}(t,s) &  =\oint_{C}\frac{d\omega}{2\pi i}\oint_{C^{\prime}}%
\frac{d\omega^{\prime}}{2\pi i}\frac{1}{\omega^{\prime}-\omega}\frac{f_{\nu
}(\omega)}{f_{\nu}(\omega^{\prime})}\frac{\left(  1+t/\Lambda\right)
^{\omega}}{\left(  1+s/\Lambda\right)  ^{\omega^{\prime}}}%
\end{align}

Here the contour $C$ encloses the poles of $f_{\nu}(\omega)$ which are located
at $\omega=0,...N$ and $C^{\prime}$encloses its zeroes, which are located at
$\omega^{\prime}=N+k-\nu;k=1,...\infty$. By taking residues at poles of
$f_{\nu}(\omega)$ we observe that $G_{\nu}(t,s)$ is an $N$-degree polynomial
in $t$ with $s$ dependent coefficients. In the same way, $S_{IM}(t)$ is an
$N$-degree polynomial in $t$, so called Jacobi polynomial. The matrix $\hat
{W}$ in $S_{IM}(t)$ remains arbitrary constant matrix, but this does not lead
to ambiguity, as it can be absorbed into definition of $Q$ by means of above
gauge transformations. In virtue of this gauge invariance we may choose
$\hat{W}=\hat{\sigma}$ in $S_{IM}(t)$.

So, we now have linear integral equation of general form%
\begin{equation}
G\left(  \hat{\sigma}+\widehat{F}\right)  Q=S
\end{equation}

\section{The $M$-limit of Pade approximant as free particle Field Theory}

The formulas of Pade approximant dramatically improve in the limit
$N\rightarrow\infty,\Lambda\rightarrow\infty$ at fixed $R^{2}=\frac{N^{2}%
}{\Lambda}$. Rather than representing numerical approximation these formulas
describe some general transformation of the whole underlying field theory,
amounting to placing it in a bag in some extra dimension. We shall refer to
this limit as meromorphic limit or $M$-limit. Note that this is
\textit{different }from naive prescription to grow $\Lambda$ linearly with $N$
which one might try first. We are moving $\Lambda\sim N^{2}$ which is much
faster. In the $M$-limit we can replace%
\begin{align}
\omega &  =N^{2}w,\\
\left(  1+t/\Lambda\right)  ^{N^{2}w} &  \rightarrow\exp(tR^{2}w),\\
f_{\nu}(N^{2}w) &  \rightarrow(-w)^{\nu-1}\exp(-\frac{1}{w}).
\end{align}
which produces the Bessel functions for $S_{IM}(t)$ . The branch of
multivalued function $(-w)^{\nu-1}$ should be chosen to positive for $w<0$ in
$f_{\nu}(N^{2}w)$ in the numerator and $w^{\prime}>0$ in $f_{\nu}%
(N^{2}w^{\prime})$ in denominator. This effectifely produces extra phase
factor $(-1)^{\nu-1}$ in $G_{\nu}(t,s)$. The overall normalization of our
kernel can be checked by its limit in case of zero dimension $\Delta_{K}^{IJ}$
where it must reduce to unity.

In the $M$-limit expansion coefficients for polynomials satisfy explicit
matrix equations (see \cite{MigdalPade}). We repeat old arguments here in what
I hope is cleaner and simpler form. We shall treat Taylor expansions as scalar
products in space of all powers of one variable:%
\begin{align}
\left\langle I\left\vert Q(t)\right\vert J\right\rangle  &
=\overrightarrow{T}(tR^{2})\cdot\left\langle I\left\vert \overrightarrow{q}%
\right\vert J\right\rangle ,\\
\left\langle I\left\vert P(t)\right\vert J\right\rangle  &
=\overrightarrow{T}(tR^{2})\cdot\left\langle I\left\vert \overrightarrow{p}%
\right\vert J\right\rangle ,\\
\overrightarrow{T}(t)  &  =\left\{  1,t,t^{2},...\right\}  ,\\
\overrightarrow{q}  &  =\{q_{0,}q_{1,}...\},\\
\overrightarrow{p}  &  =\{p_{0,}p_{1,}...\},
\end{align}

So, our matrices $q,p$ are now also vectors in space of powers of $t$ in
addition to being matrices in Hilbert space. The full dot $\cdot$ denotes
scalar products of matrices and vectors in this space of all powers of one
variable. With these notations Pade equations reduce to the following form:%

\begin{align}
\left(  \hat{\sigma}+\widehat{F}\right)  \cdot\overrightarrow{q}  &
=\overrightarrow{s},\label{MatrixVectorPade}\\
\left\langle I\left\vert s_{n}\right\vert J\right\rangle  &  =\frac
{(-1)^{n}\left\langle I\left\vert \hat{\sigma}\right\vert J\right\rangle
}{n!\Gamma(\nu_{J}+n+1)},
\end{align}

\begin{equation}
\left\langle I\left\vert F_{mn}\right\vert J\right\rangle =\frac{(-1)^{m}%
}{m!\Gamma(\nu_{I}+m)}\sum_{K}\left\langle I\left\vert g_{K}\right\vert
J\right\rangle R^{2\Delta_{K}^{IJ}}\frac{\Gamma(\nu-\Delta_{K}^{IJ}%
+n)}{(\Delta_{K}^{IJ}+m-n)\Gamma(\Delta_{K}^{IJ}-n)}.
\end{equation}

Note that this formula for matrix $F_{mn}$ has the following property. The
term with $K=0$ corresponding to the leading term in spectral density can be
recovered by setting $\Delta_{K}^{IJ}=0$ and noting that at $m\neq n$ it
vanishes due to the pole of $\Gamma(-n)$in denominator. In general case%
\begin{equation}
\frac{1}{(m-n)\Gamma(-n)}=\delta_{nm}(-1)^{m}m!
\end{equation}

Thus such term would contribute the constant $\left\langle I\left\vert
g_{0}\right\vert J\right\rangle $ to $\left\langle I\left\vert F_{mn}%
\right\vert J\right\rangle $ which is precisely the leading term $\left\langle
I\left\vert \hat{\sigma}\right\vert J\right\rangle $ \ in spectral density. We
just singled it out in our equations so that we can build perturbation
expansion. Now we se the confirmation that phase factors were chosen properly
in the kernel $G_{\nu}(t,s)$. The general solution can be represented as
matrix inversion:%

\begin{equation}
\overrightarrow{q}=\left(  \hat{\sigma}+\widehat{F}\right)  ^{-1}%
\overrightarrow{s}, \label{QInverted}%
\end{equation}
\qquad

with perturbation expansion simply corresponding to geometric series for
inverse matrix.

The relation between $P$ and $Q$ also simplifies in $M$-limit:
\begin{equation}
\overrightarrow{p}=\widehat{H}\cdot\overrightarrow{q},
\end{equation}

\begin{equation}
\left\langle I\left\vert H_{mn}\right\vert J\right\rangle =\left\langle
I\left\vert \sigma\right\vert J\right\rangle \Phi_{m}(\nu_{I}+n-m)+\sum
_{k}\left\langle I\left\vert g_{k}\right\vert J\right\rangle R^{2\Delta
_{K}^{IJ}}\Phi_{m}(\nu_{I}-\Delta_{K}^{IJ}+n-m), \label{p=Hq}%
\end{equation}

with
\begin{equation}
\Phi_{m}(a)=-\sum_{l=0}^{m}\frac{\Gamma(l+a)}{l!}%
\end{equation}
\qquad\qquad

Finally, the equation for the mass spectrum becomes operator eigenvalue
problem:%
\begin{equation}
\Psi^{\dagger}\left(  \overrightarrow{T}(m^{2}R^{2})\cdot\overrightarrow{q}%
\right)  =0; \label{SpectralEquation}%
\end{equation}

The leading CFT (or free quark) approximation corresponds to block-diagonal
$\sigma,\nu$ so that the old Bessel solution
\begin{align}
Q(t)  &  \rightarrow u^{-\frac{\nu}{2}}I_{\nu}(2\sqrt{u}),\\
P(t)  &  \rightarrow\frac{\sigma}{\sin(\pi\nu)}u^{\frac{\nu}{2}}I_{-\nu
}(2\sqrt{u}),\\
u  &  =-tR^{2}.
\end{align}
is recovered in conformal limit. After that, standard perturbation theory
wisdom can be applied, including level splitting, mixing and transitions.
Conformal classification of particles, valid in the zeroth and the first
approximation, will break in higher orders and degeneracy of conformal
multiplets will be removed. Perturbatively, there are no fundamental problems
with computation of terms, except for planar graph computation of arbitrary
conformal operators. Conformal symmetry will be broken down to the Lorentz
symmetry, and states will be classified accordingly, as relativistic
particles, by spin and internal quantum numbers.

Let us discuss the mathematical meaning of Pade regularization as described
above. First, let us compare it with conventional Pade approximation, in its
matrix form. The conventional Pade approximation at any order $N$ preserves
the Stieltjes property by placing poles at positive axis and guaranteeing
positivity of residues. The same is true with the matrix generalization
(\cite{ZinnJustin}). In perturbation expansion the Stieltjes property still
holds but it degenerates to continuous spectrum of poles reproducing powerlike
discontinuity with positive spectral density matrix. In \ this sense matrix
Pade approximation gets us closer to reality than the perturbation expansion
it approximates. It approximates the bad continuum spectrum by a good discrete
one, but with so far incorrect masses.

Or one may say that we are approximating original QCD by taking free meson
theory of infinite number of particles as an Anzatz and fitting their
parameters to approximate perturbation theory as well as possible. At finite
position $\Lambda$ the approximants of functions with continuum spectrum were
proven to converge at $N\rightarrow\infty$ absolutely in cut plane except
vicinity of the real axis where the original continuous discontinuity is
approximated by sum of delta functions.

In our case this is just the other way around. First of all, the true function
is known to be meromorphic, so that ordinary Pade approximant with finite
position $\Lambda$ will approximate meromorphic function by a finite sum of
poles. Such approximation is known to converge even faster. Second of all, we
use the $M$-limit of Pade approximant where it also becomes meromorphic. So we
approximate meromorphic function by meromorphic function, by varying pole
positions and residues to get minimal deviations from perturbation expansion
at large Euclidean momenta.

How small are these deviations? The Bessel function, corresponding to zeroth
order in perturbation expansion, approaches its powerlike asymptotics at
$t\rightarrow-\infty$ with exponential accuracy:%
\begin{align}
\frac{\sigma}{\sin(\pi\nu)}\frac{u^{\frac{\nu}{2}}I_{-\nu}(2\sqrt{u}%
)}{u^{-\frac{\nu}{2}}I_{\nu}(2\sqrt{u})}  & =\sigma u^{\nu}\left(  \frac
{1}{\sin(\pi\nu)}+\frac{2}{\pi}\frac{K_{\nu}(2\sqrt{u})}{I_{\nu}(2\sqrt{u}%
)}\right)  \\
& \rightarrow\sigma u^{\nu}\left(  \frac{1}{\sin(\pi\nu)}+2\exp(-4\sqrt
{u})+...\right)
\end{align}
Clearly, it has to be faster than any power, otherwise one would not be able
to fix the pole positions and residues. One can arrange such shifts in pole
position that the sum of poles would change only by $t^{-n}$ with arbitrary
large $n$ if we take many poles and conspire their shifts so that first $n-1$
terms of expansion in inverse powers of $t$ of these pole terms will all
vanish. However, with exponential accuracy we have in $M$-limit, there is no
room for the pole shifting. Should one forget about such requirement, one
could play with nice phenomenological "models" like $\psi(t)$ with linearly
rising spectrum of masses. We do not have a luxury of choosing the models
which fit theoretical expectations or experimental data. We derive mass
spectrum from planar graphs by regularizing them in the IR region. As QED did
in its own time, true perturbation theory must be able to remove all cutoffs
in observable quantities after certain renormalization. This is the goal of
the next Section.

\section{Dimensional regularization of LargeN QCD and Pade Theory}

In case of perturbative QCD within dimensional regularization we simply have
in above formulas%

\begin{align}
g_{k} &  =G_{k}(\epsilon)\lambda^{k},\label{dimensional_reg}\\
\mu_{k} &  =k\epsilon,\nonumber\\
d &  =4-\epsilon,\nonumber
\end{align}
where $d$\  is space-time dimension. As usual, $\lambda$ is t'Hoofts constant
in dimensional regularization, having dimension of $[m]^{\epsilon}%
$.Coefficients $G_{k}(\epsilon)$ (sum of all planar graphs of order $k$ in
coupling constant) are some calculable functions of $\epsilon$ which we need
in terms of Laurent expansion in inverse powers of $\epsilon$ starting with
$\epsilon^{-k-1}$. The limit $\epsilon\rightarrow0$ can be performed in
observables after dimensional transmutation (see later).

First of all we are pleased to note that in large order $k$ of perturbation
expansion the ratios of $\Gamma$ functions in front of $G_{k}(\epsilon)$ in
our sums over $k$ decrease as factorials of $k$ so that we have absolute
convergence of regularized perturbation expansion at any finite $\epsilon$. As
for convergence in four dimensions, it cannot be proven so easily, because it
requires taking the limit $\epsilon\rightarrow+0$ but let me make a following
simple observation. The number of planar graphs grows only exponentially, as
is well known, so that the divergence of perturbation expansion in the Large N
QCD has to do with growth of the Feynman integrals at higher order $k$. As 't
Hooft argued long ago, there are so called renormalons, namely condensing
singularities in complex plane of coupling constant $\lambda$ . These
singularities reflect precisely the discrete spectrum of masses which are
poles at $p^{2}/\mu^{2}=t_{n}$ where $\mu$ is physical mass scale. This $\mu$
behaves as $\exp(-C/\lambda)$ at small $\lambda$ so that growing mass spectrum
corresponds to singularities in $\lambda$ plane condensing to the origin as
$\frac{C}{\log(t_{n}/p^{2})}$. These singularities make the origin an
essential singularity point, eliminating any hopes for convergence of planar
graph expansion for 2-point functions. At finite $\epsilon$ the renormalon
argument still works, as in this case $\mu$ behaves as $\lambda^{\frac
{1}{\epsilon}}$ so that singularities in complex $\lambda$ plane condense to
the origin as $\left(  \frac{p^{2}}{t_{n}}\right)  ^{\frac{\epsilon}{2}}$.

But the whole point of Pade regularization is to solve this renormalon
problem. We restore the correct analytic properties of 2-point function with
infinite number of growing masses. The Pade approximant in every order of our
expansion has discrete spectrum, with density being equal to sum of delta
functions. The observables we are expanding in perturbation series, do not
depend on momentum variables-- these are precisely these discrete masses we
are expanding in powers of effective coupling constant. The renormalon
arguments simply do not apply once we made spectrum discrete. At finite
$\epsilon$ it is obvious because of extra factorial convergence we obtain in
our expansion. At $\epsilon\rightarrow0$ we cannot prove it but the common
sense says that observable quantities have no singularities at $\epsilon
\rightarrow0$ , so they should at $\epsilon=0$ be close enough to what they
are at $\epsilon=0.1$. Otherwise we will keep $\epsilon$ small and finite ,
compute observables and numerically extrapolate to $\epsilon=0$. This is of
course just a joke -- perturbation expansion allows to set $\epsilon=0$ in
every order after coupling renormalization.

One may wonder how could we have avoided the simple fact that QCD mass scale
has $\exp(-C/\lambda)$ or $\lambda^{\frac{1}{\epsilon}}$singularity at small
coupling constant. How can we expand in coupling constant at all? The answer
is simple: at finite IR cutoff $R$ there are no singularity in the mass
spectrum as a function of coupling constant. Particles are confined in a
(fifth dimension) box as part of regularization, so they have discrete rising
spectrum even without interaction. The singularities would come back in the
limit $R\rightarrow\infty$ but we take another effective coupling $\alpha$
which goes to $1$ in the IR limit without any singularities.

As physicists we all know that quarks are not very strongly interacting inside
mesons, effective coupling is small enough to make free quark picture close to
reality. It is just the limitations of modern QFT technology which prevent us
from computing masses and other observables of Large N QCD with weakly
interacting quarks at confinement scale. What we are suggesting here is the
way around these limitations of perturbative QCD.

I vividly remember discussion of my old paper with young Ed Witten in 1976 in
Boston. He quietly listened to my excited presentation of what I perceived as
a solution of the IR divergency problem in QCD and he asked only one question:
"What is a physical meaning of your infrared cutoff $R$ ?". I could not have
answered that question at that time, so I started waving hands. "Well, it
preserves the positivity and correct analytic properties of QCD and its
space-time symmetries, so there must be some physical interpretation, but
honestly I do not know it. It must be a box in some extra dimension"--said I
without any idea what I was talking about.

This was when the sales of my theory went down. Nobody needed computational
method without compelling physical picture, even if this was as fictitious as
a string in some imaginary space. This reflects the basic laws of psychology:
people care about stories more than they care about material things in life.

I still cannot answer Ed's question today, though some fantastic physical
picture have emerged with $AdS/CFT$ analogy. We now say that $R$ is a size of
the box in fifth coordinate in $Ads$ space. This picture, unfortunately, is
still incomplete. No $AdS$ model was found for large N QCD, only for some SUSY
models with degenerate set of RG equations (some coupling constants do not run
and remain as free parameters). I think that there must be more general
physical picture, without SUSY and CFT. While everybody keeps looking for this
general interpretation, we still can take $M$-limit of matrix Pade approximant
as a mathematical definition of Large N QCD and study the regularized
perturbation expansion in a hope to understand its physical meaning.

\section{Infrared regularization and effective coupling constant}

Let us consider pure QCD in glueball sector, which decouples at $N_{c}=\infty
$. The challenging problem of chiral symmetry breaking is not present here, so
this the place to start testing new perturbation expansion. All masses are
expected to be finite in this sector, so we can use any mass as a physical
scale in renormalization scheme. The most fundamental and simple quantity is
the 2-point function of stress-energy tensor.%
\begin{align}
\Pi_{\mu\nu}^{\alpha\beta}(p)  &  =\int d^{4}xe^{ipx}\left\langle \Theta
_{\mu\nu}(0),\Theta^{\alpha\beta}(x)\right\rangle ,\label{ThetaTheta}\\
\Theta_{\mu\nu}(x)  &  =\frac{1}{N_{c}}\operatorname*{Tr}(F_{\mu\alpha
}(x)F_{\nu\alpha}(x))-trace.
\end{align}

Let us consider the family of poles of Pade approximant with quantum numbers
of $\Theta_{\mu\nu}(x)$ in \ref{SpectralEquation} (so called vacuum Regge
trajectory) :%
\begin{equation}
m_{n}^{2}=R^{-2}\exp\left(  f_{n}\left(  \lambda_{R}\right)  \right)
\label{lowestMass}%
\end{equation}

where the function $f_{n}\left(  \lambda\right)  $ can be obtained
perturbatively from above matrix equation for $Q(t)$. We denote the running
coupling constant at scale $R^{-1}$ as $\lambda_{R}$, assuming that the UV
regularization is already removed, so that $\epsilon=0$.

Let us look at the structure of the relation for $m_{n}^{2}$ . There is one
important fact in Pade theory: it always overestimates the masses. In other
words, at any fixed $\Lambda$ every mass monotonously decreases with $N$.
Taking the liberty of assuming that the same property holds in the $M$-limit,
where $R=N/\sqrt{\Lambda}$ remains finite after setting $\Lambda=\infty$, we
conclude that $\log(m_{n}^{2})$ monotonously decreases with scale $\log
(R^{2})$. In terms of $\beta$-function this inequality reads:%
\begin{equation}
-\beta(\lambda)f_{n}^{^{\prime}}\left(  \lambda\right)  <1
\end{equation}

At small $R$ this decrease is trivial, as the factor $R^{-2}$ is the fastest
changing factor in perturbative region. The masses of free particles in a box
size $R$ decrease as $1/R$ when the box grows. However, we expect that with
increase of $R$ the growth of effective coupling constant $\lambda_{R}$ will
stop this decrease so that asymptotically at large $R$ the masses approach
finite limits from above. In that limit the above inequality turns into
equality%
\begin{equation}
-\beta(\lambda)f_{n}^{^{\prime}}\left(  \lambda\right)  \rightarrow1^{-}.
\end{equation}

In terms of function $f_{n}\left(  \lambda_{R}\right)  $ this mean that it
goes to $+\infty$ as $\log(R^{2})$. In order to interpolate between weak and
strong coupling regions let us generalize this formula by applying Legendre
transform:%
\begin{align}
\log(\frac{m_{n}^{2}}{\mu^{2}})  &  =\lim_{\alpha\rightarrow1^{-}}\Phi
_{n}(\alpha),\label{Legendre}\\
\Phi_{n}(\alpha)  &  =\min_{R}\left(  f_{n}\left(  \lambda_{R}\right)
-\alpha^{2}\log(\mu^{2}R^{2})\right)  ,\\
\Phi_{n}^{^{\prime}}(\alpha)  &  =-2\alpha\log(\mu^{2}R^{2}).
\end{align}

At any $\alpha<1$ the RHS of above formula goes to $+\infty$ as $(1-\alpha
^{2})\log(R^{2})$ at $R\rightarrow\infty$ . On the other hand, at small $R $
it also goes to $+\infty$ as $-\alpha^{2}\log(R^{2})$. Therefore there must be
at least one minimum in between. In virtue of the Pade theorems about
monotonic mass decrease we may expect that there is only one minimum. The idea
behind the $\alpha$ expansion we proposed in the old paper (\cite{MigdalAlpha}%
) is that there is smooth interpolation between perturbative region of small
$\alpha$ and confining theory at $\alpha=1$.

The equation for the minimum in \ref{Legendre} involves the $\beta$-function:%
\begin{equation}
-\beta(\lambda)f_{n}^{^{\prime}}\left(  \lambda\right)  =\alpha^{2}.
\label{MaxEq}%
\end{equation}

As the $\beta$-function starts with $-a\lambda^{2}$ with positive $a$ we get:%
\begin{equation}
\lambda_{n}(\alpha)\rightarrow\frac{\alpha}{\sqrt{af_{n}^{^{\prime}}(0)}}.
\end{equation}

So, in general, the Legendre transform $\Phi_{n}(\alpha)$ starts with some
constant at $\alpha=0$ then grows, reaches its maximum at some point where
$\mu R=1$, then starts decreasing and reaches the limit at $\alpha=1$. This is
much smoother behavior than the $R$-dependence of original function
$f_{n}\left(  \lambda_{R}\right)  -\log(R^{2})$ , which starts at $+\infty$ as
$-\log(R^{2})$ at small $R$ then decreases and reaches the limit at
$R\rightarrow\infty$. This was the purpose of the Legendre transform to
achieve dimensional transmutation and obtain smoother behavior. No finite
order of perturbation expansion could have reproduced the confining
asymptotics at $R\rightarrow\infty$. On the contrary, within the $\alpha
$-expansion starting with the second order, we can have expected properties:
linear growth at small $\alpha$ then maximum at finite $\alpha$ then descend
to another limit at $\alpha=1$. The drama was totally eliminated from
confinement story by means of this Legendre transform.

The first 4 coefficients of $\alpha$-expansion for all basic trajectories
(vector, scalar, vacuum) were computed in my old paper (\cite{MigdalPade}).
For the operator with $n-2$ extra derivatives%
\[
\Theta_{\mu\nu...}^{(n-2)}(x)=\frac{1}{N_{c}}\operatorname*{Tr}(F_{\mu\alpha
}(x)\nabla...\nabla F_{\nu\alpha}(x))-traces.
\]

the coefficients $f_{n}^{^{\prime}}(0)$ are proportional to anomalous
dimensions in leading order%
\begin{equation}
\gamma_{n}^{^{\prime}}=\frac{6}{11}\left[  \frac{1}{3}-\frac{4}{n(n-1)}%
-\frac{4}{(n+1)(n+2)}+4\sum_{j=2}^{n}\frac{1}{j}\right]
\end{equation}

which is positive for $n>2$ and vanish at $n=2$ in virtue of conservation of
energy-momentum tensor. This means that lowest mass cannot be taken as a
definition of physical scale in this procedure. We can take the next one,
\thinspace$n=3$, use the relation between effective coupling and $\alpha$ and
compute remaining masses in terms of this particular $\alpha$ corresponding to
\thinspace$n=3.$\qquad%

\begin{align}
\log\left(  \frac{m_{n}^{2}}{m_{3}^{3}}\right)   &  =f_{n}(\lambda
)-f_{3}(\lambda),\\
-\beta(\lambda)f_{3}^{^{\prime}}\left(  \lambda\right)   &  =\alpha^{2}.
\end{align}
\qquad

This expansion is constructed so that all terms are universal and calculable,
with zeroth term having good resemblance to reality, as we know from the old
paper as well as its rebirthing with $AdS/CFT$ models. One may argue that
there are infinitely many ways to build such an expansion and we agree with
that. But the same can be said about Wilson's $\epsilon$ expansion, which
interpolated between $4$ and $4-\epsilon=3$ dimensions in the phase transition
theory. There are many ways of analytic continuation into fractional dimension
of space, but the most natural one chosen by Wilson turned out to work well in
practice. It remains to be seen whether this $\alpha$ expansion will be as lucky.

\section{Conclusion}

The approach we revived in this paper is opposite to modern quest for String
Theory solution of Large N QCD. Instead of finding peculiar string theory
equivalent to perturbative QCD in the UV region we postulate existence of such
string theory and are studying its properties without a luxury of some local
2D field theory as a dual definition. We rather introduce most general free
meson theory in 4D with unknown masses and coupling constants to QCD gauge
invariant composite fields and derive explicit equations for these masses and
coupling constants.

The underlying idea is that perturbative QCD is very close to reality, as the
effective coupling constant at hadron scale is small. It is just the IR
divergencies which prevent us from perturbative calculations of mass ratios.
Once we regularize the perturbation theory we can expect rapid convergence,
because the planar graphs grow only a power of order of perturbation
expansion. Renormalons look like a paper tiger: they disappear after
regularization. On top of general analytic arguments there are extra
factorials in denominator of our expansion, arising from transformation from
perturbative QCD to the mass spectral operator.

The educated reader (of older generation) may say: wait a minute, but where
are SVZ vacuum condensates? My first answer: they are no longer needed as we
do not truncate the perturbation expansion. We rather argue that this
expansion converge. SVZ condensates were designed to phenomenologically
describe renormalons: high momentum regions in Feynman integrals of higher
order, behaving as powers of QCD mass scale. These powerlike terms balanced
the decrease of masses as function of $R$ and SVZ obtained good agreement with
experiment by matching one-loop QCD with sum of pole terms at the scale
determined by the vacuum condensate terms.

On the other hand, in case the SVZ condensates would be derived and computed
from the first principles, there is no problem adding them to the expansion,
as they are just another powerlike corrections, which we already considered.
This problem should be solved beyond the scope of planar graphs, at
non-perturbative level. The nonperturbative formulation of the Large N QCD is
given by the loop equation. My last study of this equation \cite{MigdalHidden}
led to algebraic formulation, with the string position operator $X_{\mu}$
playing the role of Witten's master field in dual space (for momentum Wilson
loop) and satisfying some nonlinear algebraic relations. This formulation is
suited for the expansion in powers of momenta and it fits the meromorphic
structure presented here. Our next challenge would be to match these two
approaches and obtain mass spectrum equation relating $Q$ to operator $X_{\mu
}$. It is tempting to speculate that VEV's of powers of $X_{\mu}$ are somehow
related to the SVZ condensates. 

One last word about notorious Pade approximation. I hope the reader
understands by now that there are no approximations involved in our theory.
This is regularization rather than approximation. We impose correct analytic
properties and symmetries of large N QCD beyond perturbation theory by first
using Matrix Pade approximation in Hilbert space and then taking $M$ limit
when approximation becomes realistic in a sense that it has all the desired
properties including infinite mass spectrum.

One way of viewing this regularization is to say that CFT is defined up to
terms exponential in momentum in the UV region. All the powerlike terms come
from VEV in OPE, but possible exponential terms represent nonperturbative
effects. Our Pade regularization introduces such terms in a unique way,
preserving unitarity and analiticity. So, for any $R$ we may declare that the
theory is defined beyond perturbation expansion. The limit $R\rightarrow
\infty$ brings us back to "original" theory whatever it is. In case there is
confinement, the mass spectrum will stay finit at $R\rightarrow\infty$,
otherwise it will condense to contimuum of freely flying gluons and quarks.
So, within our formulation, confinement comes about as a result of conformal
symmetry breaking by SVZ condensates of our CFT.

Recently, my old work was noticed and misinterpreted by several authors. Their
conclusion was that there were many phenomenological formulas with poles
approximating logarithm, say, $\Psi$ function. I certainly agree with this
statement, but it has nothing to do with the theory developed in that work.

There is no systematic treatment of large $N$ QCD as a theory of infinite
number of free particles with \ masses given by poles of $\Psi$ functions.
Such theory must start in Hilbert space to take into acount whole amount of
unitarity+ analiticity requirements. One cannot treat various channels
separately. As for the Bessel functions, they naturaly arise in zeroth
approximation of CFT because that this sum of pole terms approaches conformal
limit as fast as possible in complex plane (as exponential of momentum).

However, by itself the ratio of Bessel functions does not represent any
consistent field theory. It contains artificial cutoff $R$ which must tend to
infinity after summation of perturbation expansion. If we limit ourself to the
ratio of Bessel functions and tend $R$ to infinity we recover original CFT. It
is only after dimensional transmutation and taking the limit $\alpha
\rightarrow1$ we remove ambiguity of initial approximation and come back to
real QCD.

Note that conformal symmetry of the leading order was absolutely essential to
obtain calculable perturbation expansion. It is because of conformal symmetry
our leading order $Q$ matrix becomes block-diagonal in space of conformal
tensors, so that we can invert it in Pade equations. The conformal group
representation of meson states is broken down to general Lorentz group
representation in the second order in our perturbation expansion.

The interpolation to physical limit when the approximation disappears
completely is similar to UV regularization of QED half century ago. The cutoff
enters only in logarithms, effectively renormalizing running coupling
constant. By renormalization, using Legendre transform, we trade \ this cutoff
for the effective coupling constant $\alpha$ which should be set to $1$ in the end.

I no longer hope that anyone will pick this theory and actually carry out
these computations of Meson spectrum. I will do my best to finish this work
myself. The terms of this renormalized perturbation expansion are all
calculable, and we presented explicit formulas in this paper. Maybe these
formulas will convince the authors of recent papers on the subject that soft
corrections to CFT \textbf{do} lead to corrections to mass spectrum: one of
the goals of the present paper was to clarify this misunderstanding.

The hardest problem is to compute ordinary planar graphs for 2-point functions
with dimensional regularization. I am aware of 3 and 4 loop calculations, but
this may not be enough. Maybe some recent progress in perturbative
calculations in SUSY Yang Mills theory can help here?

\bigskip

\section{Acknowledgements}

\bigskip

This paper was presented at A.B.Migdal100 Conference in Landau Institute in
June 2011, and I am grateful to Sasha Dugaev, Sasha Belavin, and other
participants for  support and discussions of the paper.

I warmly thank Ilya Khrzhanovsky, Dau and Dima Kaledin, the Director of the
Institute (Moscow, USSR), for their overwhelming hospitality during July 1956,
and participants of Soviet-American Conference: David Gross, Igor Klebanov and
Nikita Nekrasov for their comments and suggestions, but mostly for their encouragement.

This paper was also presented at Carg\`{e}se PhyMSI conference in July 2011,
and I would like to thank Volodya Kazakov and Ivan Kostov for their
hospitality under pressure. On scientific side, I greatly benefitted from
discussions with Ed Witten, Pasha Wiegman and Andrej Okunkov in Carg\`{e}se.

Thank you, old and new friends, you made me feel young again.

\end{document}